\documentclass[10pt,journal,compsoc]{IEEEtran}

\usepackage{ctable}
\usepackage{caption}

\ifCLASSOPTIONcompsoc
  \usepackage[nocompress]{cite}
\else
  \usepackage{cite}
\fi

\usepackage{graphicx}
\usepackage{float}
\usepackage{amssymb,amsfonts,amsmath,amsthm}

\usepackage{breqn}
\usepackage{algorithm,algpseudocode}
\usepackage{enumitem}
\usepackage[hidelinks]{hyperref}
\usepackage{multirow}
\begin{document}

\title{A Sequence-Aware Recommendation Method Based on Complex Networks}

\author{Abdullah~Alhadlaq, Said~Kerrache, Hatim~Aboalsamh
\IEEEcompsocitemizethanks{\IEEEcompsocthanksitem King Saud University, College of Computer and Information Sciences, Riyadh, 11543, KSA.\protect\\
E-mail: 436106753@student.ksu.edu.sa}
}

\IEEEtitleabstractindextext{
\begin{abstract}
 Online stores and service providers rely heavily on recommendation softwares to guide users through the vast amount of available products. Consequently, the field of recommender systems has attracted increased attention from the industry and academia alike, but despite this joint effort, the field still faces several challenges. For instance, most existing work models the recommendation problem as a matrix completion problem to predict the user preference for an item. This abstraction prevents the system from utilizing the rich information from the ordered sequence of user actions logged in online sessions. To address this limitation, researchers have recently developed a promising new breed of algorithms called sequence-aware recommender systems to predict the user's next action by utilizing the time series composed of the sequence of actions in an ongoing user session. This paper proposes a novel sequence-aware recommendation approach based on a complex network generated by the hidden metric space model, which combines node similarity and popularity to generate links. We build a network model from data and then use it to predict the user's subsequent actions. The network model provides an additional source of information that improves the accuracy of the recommendations. The proposed method is implemented and tested experimentally on a large dataset. The results prove that the proposed approach performs better than state-of-the-art recommendation methods. 
\end{abstract}

}

\maketitle

\IEEEdisplaynontitleabstractindextext

\IEEEpeerreviewmaketitle

\IEEEraisesectionheading{\section{Introduction}\label{sec:introduction}}
	A recommendation system (RS) is a software tool that uses various techniques and algorithms to filter the relevant information from the vast information found in an online platform based on multiple factors to provide users with item recommendations that most likely suit their preferences. RSs have been applied in various domains, including travel and hotel industries, online shopping, books, and movie recommendations \cite{Bobadilla2013}.

Conventionally, the recommendation problem is abstracted as a matrix-completion problem where users correspond to rows, items correspond to columns, and the numerical cell values indicate the user-item ratings. The goal of matrix completion is to predict the ratings of unseen items for a given user based on historical data \cite{jannach2016recommender}. Although this abstraction has proved helpful in various ways, it suffers from the limitation of not utilizing the sequence of user interaction logs that are often available in real applications \cite{quadrana2018sequence}. Moreover, conventional recommendation systems assume user profile availability and long-term historical data. However, such long-term data does not always exist for many reasons, such as the user being new to the system, having opted not to log in, or the user is enabling tracking countermeasures \cite{jannach2018keynote,fang2019deep}. For this, sequence-aware recommendation systems (SARS) have recently been developed to harness the rich information from logged users' interactions with the system. The goal is to derive predictions for subsequent user actions based on the recent series of actions in the ongoing user session, thus, bringing highly relevant and practical computational tasks to real-life applications.

Despite the ongoing efforts to improve SARS, the accuracy of recommendations remains an open challenge. Most of the existing complex models, including deep learning and matrix factorization, are outperformed by straightforward trivial approaches such as the $k$ nearest neighbors' approach \cite{ludewig2018evaluation, latifi2021session}. One of the most promising directions to remedy this is using graph (or network) models to generate recommendations. In network-based recommenders, where nodes represent users or items and the weighted links between the nodes represent relevancy, the analysis derived from the graph structure can produce accurate predictions. Many network-based recommendation approaches have been introduced in the literature, including approaches based on real-world networks \cite{huang2007analyzing, zhou2007bipartite, zanin2008complex}. However, most approaches use networks without underlying models or established properties. Using complex network models with well-understood and proven properties can improve the quality of recommendations and constitutes a promising research direction. 

One of the most relevant recent advances in complex network modeling to the field of sequence-aware recommendation is the development of efficient navigation and routing algorithms \cite{ boguna2010sustaining,boguna2021network}. Since the interaction between the user and items in SARS can be seen as navigation in the space of items, this hints at the possibility of using complex network models to guide the recommendation process. 
In this paper, we propose a novel sequence-aware recommender system approach that takes advantage of complex network models, primarily the hidden metric space model \cite{serrano2008HMS}, to generate more accurate recommendations. The hidden metric space model is a complex network model that suggests the existence of a hidden metric space underlying any observed complex network. The distance between the nodes abstracts their similarities, whereas their degrees represent popularity \cite{boguna2021network}.
Our method utilizes the rich navigation information within the large logs of sequentially ordered actions to improve recommendation accuracy. The resulting algorithm is trained and tested on a large public dataset and evaluated using standard performance measures to compare its performance to state-of-art methods. 	

\begin{figure*}[!t]
	\centering
	\includegraphics[scale=0.33]{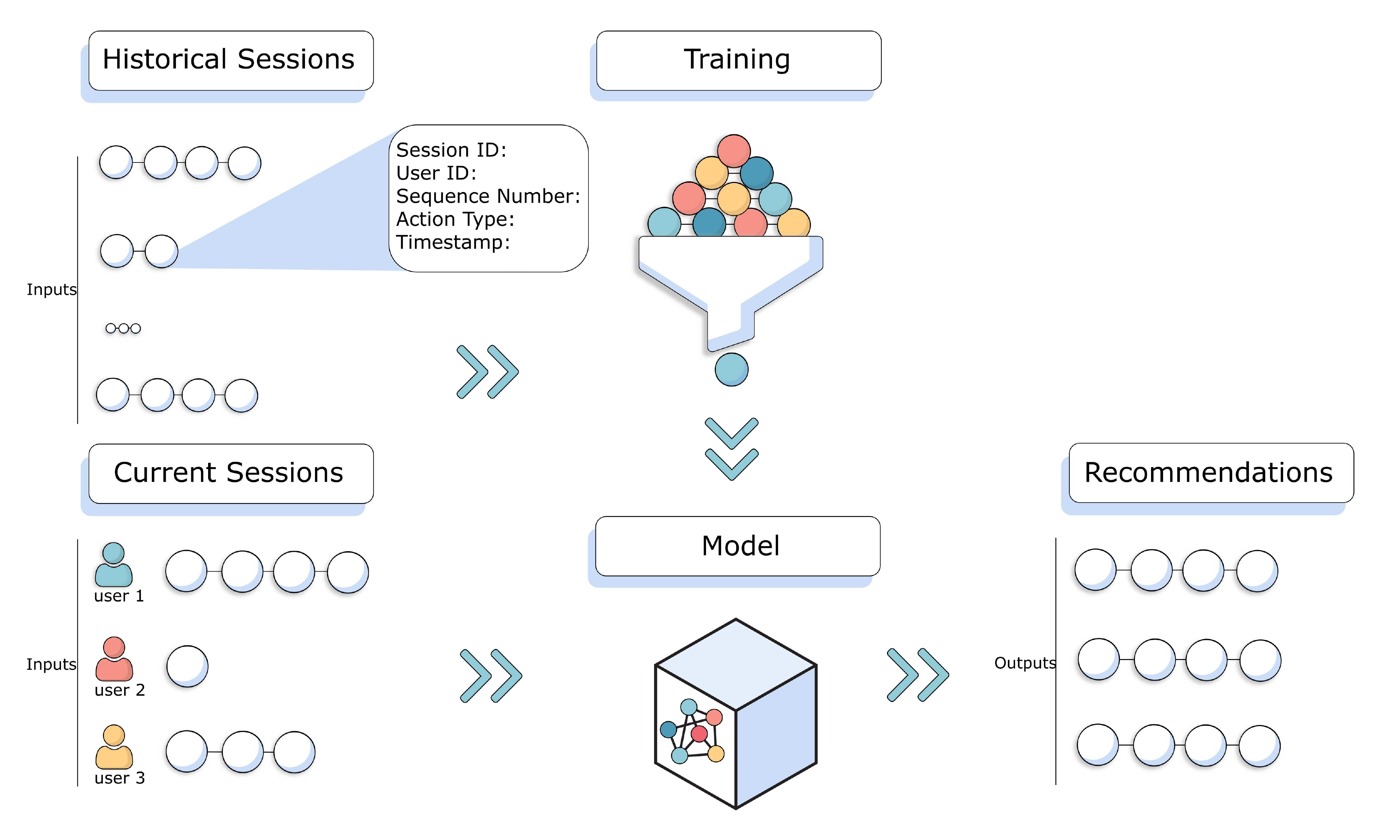}
	\caption{A high-level representation of sequence-aware recommender systems process.}
	\label{fig:SARS}
\end{figure*}

The remainder of this paper is organized as follows. Section \ref{sec:background} covers background and related work. Section \ref{sec:proposed} describes the details of the proposed method. Section \ref{sec:exp} presents the experimental setup and performance evaluation results. Finally, Section \ref{sec:conclusion} summarizes the work and presents future research directions.

\section{Background and Related Work} 
\label{sec:background}
In this section, we briefly introduce recommender systems, particulary sequence-aware recommender systems, followed by an overview of the complex networks and the hidden metric space model.

\subsection{Sequence-aware Recommender Systems}
The goal of sequence-aware recommender systems is to predict the evolution of the user's current session. They achieve this goal by predicting the user's next action using extracted features from historical sessions' logs tracked by the application \cite{jannach2018keynote }. Fig. \ref{fig:SARS} shows a schematic illustration of the recommendation process under the sequence-aware setting. The recommender system's input takes the form of an ordered set of users' logged actions (sessions). These actions consist of various user interactions towards an item such as 'clicks', 'views', or 'purchases', or actions towards the application such as 'searches' or 'applies filters'. The computational task of sequence-aware recommenders is mainly to process the input to build a model that attempts to identify recurring patterns in the sequences of actions. For example, this pattern may reflect the co-occurrence of actions or their sequential ordering. Finally, the system's output is the predictions for the user's next action (or set of actions) derived from the identified pattern based on the current user session.

\begin{figure}[!t]
	\centering
	\includegraphics{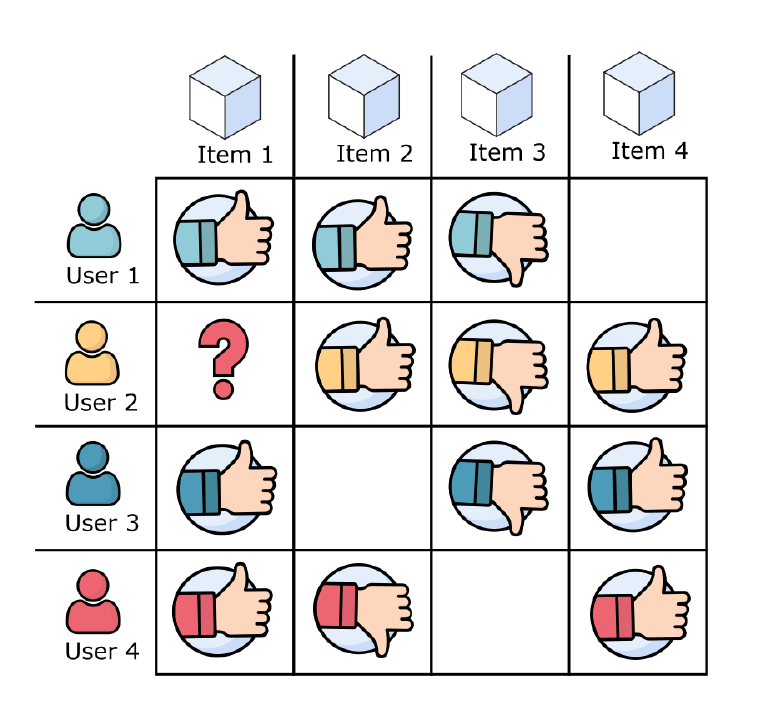}
	\caption{Representation of the user-item matrix in traditional recommender systems. The goal of the recommender is to complete this matrix.}
	\label{fig:RS}
\end{figure}

Although the output from traditional RS and SARS aims at the same goal: to provide an item recommendation for the users, their settings and characteristics are distinct in several ways. First, the standard input for traditional RS takes the form of user, item, and rating tuples without any information about the user's behavior or interactions with the items or the application. In contrast, the primary input for SARS is the rich sequential session data. Moreover, a standard RS takes the form of a matrix-completion problem, as illustrated in Fig. \ref{fig:RS}, in which the task is to fill the predicted ratings given by users to items based solely on the long-term observed user behavior and preferences. This formulation, however, can not accommodate the sequential input representation \cite{jannach2016recommender}.

The problem of sequence-aware recommendation has been increasingly attracting researchers. Early research efforts proposed oversimplified model-free attempts to provide recommendations driven by data pattern mining and association rules techniques. However, since the 2010s, machine learning and time-series techniques have shifted the algorithmic approaches towards adopting more sophisticated models to tackle the problem \cite{wang2021survey}. Models such as Markov chain models \cite{rendle2010factorizing,he2016fusing}, graph-based models \cite{xiang2010temporal,yuan2014graph,li2021sequence}, factorization-based models \cite{kabbur2013fism}, recurrent neural networks \cite{jannach2017recurrent,hidasi2015session,tan2016improved} and attention mechanisms \cite{xu2021long} contributed to improving sequence-aware recommendation systems \cite{ludewig2018evaluation}. 

Several graph-based approaches for sequence-aware recommenders exist in the literature that extends the work proposed for traditional recommender systems. In these methods, the graph is often modeled as a bipartite graph having items and users on separate sides. However, this poses a limitation as these approaches fail to exploit the rich information in sessions. To overcome this issue, authors in \cite{xiang2010temporal} proposed the session-based temporal graph (STG) method that incorporates temporal dynamics and short- and long-term preferences. The STG graph is represented as a two-sided bipartite graph (tripartite), with one side of the graph connecting users with items they have interacted with before (i.e., long-term interactions) and the other side connecting the current user session with session items (i.e., short-term interactions). The graph is then traversed to recommend items via random walk. 

Similarly, the Geographical-Temporal influences Aware Graph (GTAG) method \cite{yuan2014graph} uses a tripartite graph representation. However, GTAG incorporates geographical location to enhance the point of interest recommendations and traverse the resulting graph model in a breadth-first propagation manner. Both methods, however, as in all multipartite graphs models, which do not allow edges between nodes on the same side, fail to exploit information on any independent side, for example, item-to-item data. Additionally, both methods construct a simple graph without an underlying model. Conversely, the proposed method in this article relies on constructing a graph with an underlying model and established properties to guide the recommendations.

In recent years, researchers have attempted to achieve accurate recommendations by providing explicit interpretations of the challenges and tasks in the SARS and then attempting to tackle the identified challenges by designing appropriate algorithms. 
Quadrana et al. \cite{quadrana2018sequence} suggest that the main tasks of sequence-aware recommender systems are adopting the recommendation context, such as the current weather or the current user location, detecting the shared and individual trends, identifying the repeated patterns in user behavior, and finally identifying the order constraints of the sequence in the session. 
Jannach et al. \cite{jannach2017session} state the price discounts and current offers on items as a success factor for SARS and emphasize the impact of the user’s short-term intentions rather than only long-term ones. Wang et al. \cite{wang2021survey} further focus on understanding the inner interactions within the session and the interactions between the sessions as the two key challenges to reducing the complexity of SARS structure.

\subsection{Complex Networks and Hidden Metric Space Model}
Complex networks refer to a class of graphs that exhibit nontrivial topological features that can be observed in real-life networked systems. Those features are not observable in simple graphs such as regular and purely random graphs but in networks resulting from complex natural phenomena such as social, biological, technological, and physical systems \cite{wang2003complex}. For example, when modeling a real-life social network, where the nodes represent the people and the edges between the nodes represent the relationships between them, the resulting graph possesses a complex network structure. Its topology reveals some fundamental complex network properties and characteristics \cite{yuan2010small-worldness}, such as the small-world \cite{WSmodel998} and the scale-free properties. Scale-free networks gained significant interest within the network science community upon the publication of the Barabási and Albert model \cite{barabasi2000scale}. The latter generates scale-free networks with short paths and highly connected nodes as hubs and is considered one of the earliest models to capture real-life natural network properties \cite{barabasi2009decade}.

Efforts to design network models that capture more faithfully the properties of real-life networks have continued after the work of Barabási and Albert model \cite{barabasi2000scale}.
The hidden metric space model \cite{serrano2008HMS} suggests the existence of a hidden metric space underlying any scale-free complex network and contains all the nodes of the observed network. The nodes are positioned and linked in the hidden metric space based on their similarity and popularity. The distance between nodes abstracts the nodes' similarities, and their degrees reflect their popularity. Thus, the shorter the distance between nodes in the hidden metric space and the more popular they are, the higher probability of them connecting in the observed network.

One of the successful applications of the hidden metric space model is its capacity to guide the routing function in networks by greedily moving toward the node nearest to the target \cite{boguna2009navigabilityHMS}. Unlike many complex network models, such as the small-world model or Barabási and Albert model, the hidden metric space model generates an embedding of the network, which provides a useful representation of the items for downstream applications. Furthermore, the model not only aims at capturing the network's structural properties but also describes how these properties are connected with essential functionalities. The successful use of the model in various domains, including developing efficient information routing algorithms \cite{boguna2010sustaining}, network modeling \cite{boguna2021network}, information routing and signaling \cite{boguna2009navigabilityHMS}, link prediction \cite{papadopoulos2014network, kerrache16, kerrache20} and recommender systems \cite{alhadlaq22} hints at the possibility of applying it to guide the recommendation in a SARS settings.

\begin{figure*}[!t]
	\centering
	\includegraphics[scale=0.50]{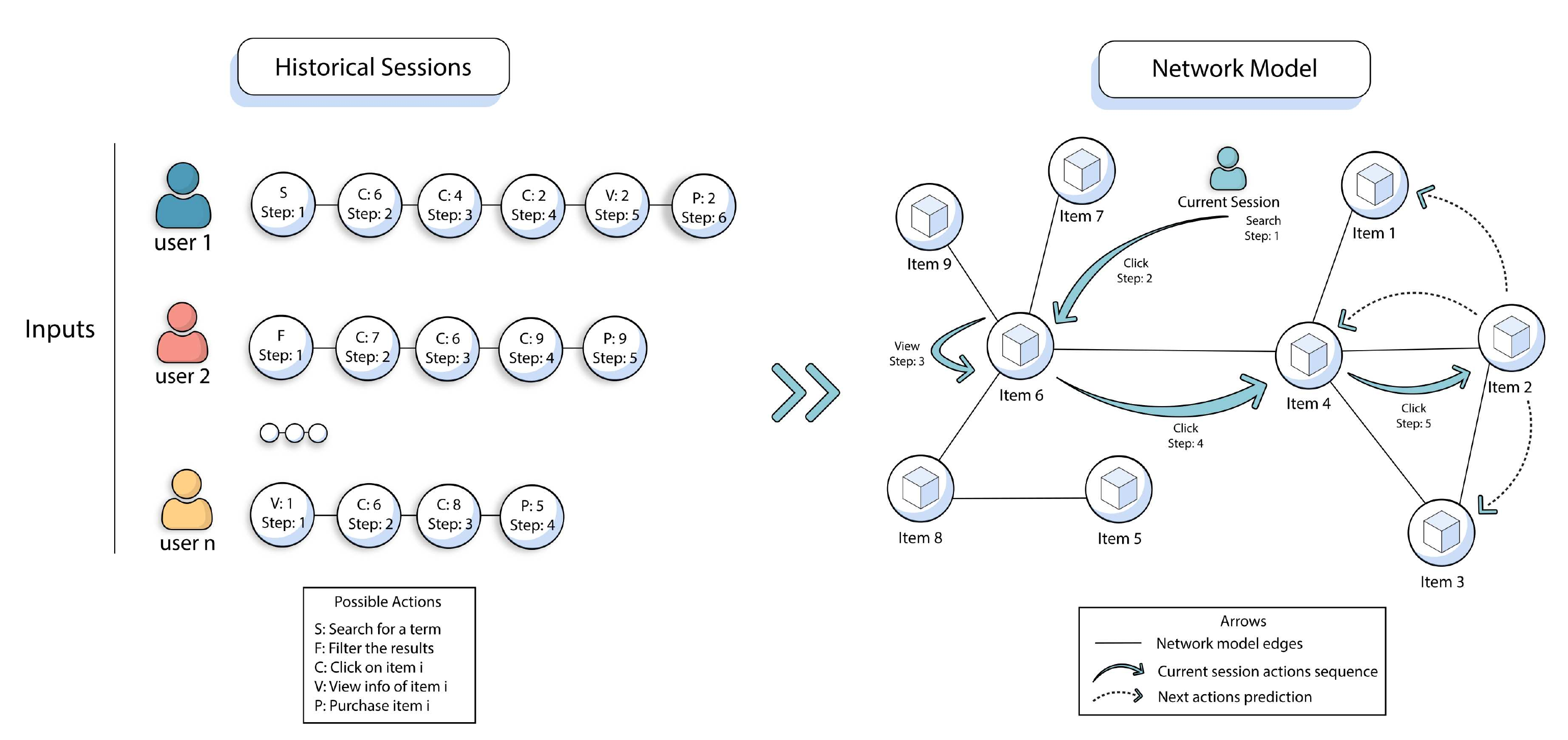}
	\caption{Illustration of the proposed approach. Items are modeled as nodes in a complex network and are connected based on similarity and popularity.}
	\label{fig:Proposed}
\end{figure*}

\section{Proposed Method}
\label{sec:proposed}

In this paper, we propose a sequence-aware recommendation method based on complex networks where items are modeled as a graph embedded in $D$-dimensional space with a scale-free network structure controlled by an underlying hidden metric space. Since the goal of a SARS is to predict the next click the user will perform in a session, the problem is defined as the task of ranking the most probable next item. The output of the recommender is an ordered list of items $r \in R^*$ for each session $s$, where $R^*$ is a subset of the group of all possible items permutations, and each predicted ordered list of recommendations $r$ consists of $i$ items of length between $0$ and $t$. First, each item in the system $i$ is assigned a point $x^i$ in a $D$-dimensional Euclidean space: 
\begin{equation}
	x^i=\left(x^i_1, x^i_2, \ldots x^i_D\right)^T , i = 1,\ldots n,
\end{equation}
where $n$ denotes the number of items. The position of the items reflects their description or features. Hence, each dimension represents a latent feature or a combination of features relevant to the item. The probability $p_{ij}$ of connection between two items nodes $i$ and $j$  is based on two factors, their similarity reflected by their distance and popularity (or degree) as derived from the hidden metric model:
\begin{equation} 
	\label{eq:HMSM}
	{p}_{ij}={\left(1+\dfrac{d^2(x^i,x^j)}{{\kappa_i\kappa_j}}\right)}^{-\alpha}, 
\end{equation}
where $\kappa_i$ and $\kappa_j$ denotes the hidden degree (popularity) of items $i$ and $j$, respectively, and $d^2(x^i,x^j)$ is the squared distance between them. The parameter $\alpha > 1 $ controls the weights given to distance and popularity. 

According to Eq. \eqref{eq:HMSM}, the connection probability between nodes in the generated network increases when popularity increases and decreases with the squared hidden distance (dissimilarity) increase. The reason behind this choice is that generated network will have three characteristics observed in real-life complex networks. First, a pair of popular nodes will have a higher probability of being connected even if they are far away from each other or dissimilar. A pair of nodes with moderate average popularity will be connected if the distance is moderate, and a pair of nodes with low popularity on average will get connected if the distance is short.  

As illustrated in Fig. \ref{fig:Proposed}, items are embedded in a $D$-dimensional space, where the distance between them encodes their similarity, which, combined with item popularity, results in a network model. The model can be used to compute the probability of a pair of nodes being connected and predict the next clicked item in a sequence.
We need to calculate the item's popularity and similarity to construct the model. The node's degree represents its popularity and can therefore be directly calculated from the number of clicks or interactions the item has in historical sessions. Next, we assign each node a position in $D$-dimensional space to calculate the distances. Since we cannot readily obtain the positions of the nodes, we infer them from the history of interaction sessions. We first estimate the link probabilities directly from historical data and then use them to infer the items' positions. There are arguably many ways to estimate connection probabilities. However, the method we adopt in this work is to assume that the connection probability is proportional to the number of items co-occurrence in sessions calculated by the cosine similarity as follows:

\begin{equation} 
	\label{eq:sim}
	{p}_{ij} \propto sim(i,j) = 
	\frac{\displaystyle\sum_{s\in S}i_s j_s}
	{\sqrt{\displaystyle\sum_{s\in S}{i_s^2}}
		{\sqrt{ \displaystyle\sum_{s\in S}{j_s^2}}}},
\end{equation}
where $i_s$ is set to $1$ if item $i$ is present in session $s$ and $0$ otherwise. From that, we obtain the set $P$ of all connections probabilities $p_{ij}$ such that $p_{ij}>0$. It is now possible to deduce the distance $d^2_{ij}$ in Eq. \eqref{eq:HMSM} as follows: 
\begin{equation}
	\label{eq:dist}
	d^2_{ij} =\kappa_i \kappa_j  \left(p_{ij}^{-1/\alpha}-1\right).
\end{equation}

Hence, the problem of estimating the items' positions is reduced to finding coordinates $x^i$, such that the resulting Euclidean distances are as close as possible to the distances prescribed by Eq. \eqref{eq:dist}. This problem falls into the category of nonlinear optimization problems that can be solved by minimizing an objective function that measures the difference between the computed and observed distances:

\begin{equation} 
	\label{eq:f}
	\sum_{p_{ij} \in P} \left( \left\|x^i - x^j\right\|_2^2- d_{ij}^2\right)^2,
\end{equation}
where $\|\cdot\|_2$ stands for the $L_2$ norm.
In order to avoid overfitting, a regularization term is added, resulting in the following objective function:
\begin{multline} 
	\label{eq:ff}
	f(x^1, \ldots, x^n)= \sum_{p_{ij} \in P} \left( \left\|x^i - x^j\right\|_2^2- d_{ij}^2\right)^2\\
	+ \lambda \sum_{i=1}^n \left\|x^i\right\|_2^2.
\end{multline}
Since this objective function is not convex, only local minima can be computed. Several algorithms for local optimization exist to solve this minimization problem, and most of them require computing the gradient of the objective function: 
\begin{equation}
	\label{eq:gfi}
	\frac{\partial f}{\partial x^i} = \sum_{p_{ij},  p_{ji} \in P} 4 \left(x^i - x^j\right) \left( \left\|x^i - x^j\right\|_2^2- d_{ij}^2\right) + 2 \lambda x^i.
\end{equation}

The proposed method's objective is to predict a list of items ordered by the likelihood that the user will click next in the active session. Once we have the item embeddings, we can use different techniques to capture the user's short-term interest in the current session to predict the next item. We propose to use the $K$ nearest neighbors technique, where distance is measured in terms of connection probability. Thus, the method recommends the items with the highest connection probability to the most popular item the user interacted with in the active session. In other words, the method generates a list of $t$ items ordered decreasingly by the probability of connection to the most popular item in the active session as calculated by Eq. \eqref{eq:HMSM}.

\section{Experimental Evaluation}
\label{sec:exp}
This section presents the experimental framework used to evaluate the proposed method's performance. We present the experimental setup and describe the dataset, the dataset preparation techniques, the competing methods, the evaluation metrics, the training method, and the parameter tuning technique. Finally, we compare the overall performance results against competing methods.

\subsection{Dataset Description}
The dataset used in this paper is the Trivago dataset \cite{adamczak2020session, knees2019recsys} presented in the 2019 RecSys challenge obtainable online from a publicly available data source (\url{https://recsys2019data.trivago.com}). The dataset contains sequences of various users actions on the Trivago hotels booking website. It contains 981,655 users, 927,142 hotels, and 1,202,064 sessions containing a total of 19,715,327 user actions.

The dataset is pre-split into train and test sets, the splits were used as they are in this experiment. We cleaned the dataset by removing sessions that do not lead to a hotel booking. Each session in the test set ends with a hidden hotel booking action and a list of a maximum of 25 hotels in the impressions list, and the goal is to reorder the list of impressions in decreasing order by the likelihood of user clicks.

\subsection{Evaluation Criteria}
The proposed approach is evaluated using the test set by applying standard metrics for the sequence-aware recommender systems evaluation. There are different evaluation techniques and accuracy measures for sequence-aware recommender methods in the literature. However, since the output of a sequence-aware recommender system typically takes the form of ordered lists, it is applicable to use standard information retrieval accuracy metrics \cite{ludewig2018evaluation} such as mean reciprocal rank (MRR) and mean average precision (MAP). 

The MRR measures the place of the correct item in the predicted list. Given a test set $S^{Test}$ containing previously unseen sessions, the MRR is defined as:
\begin{equation}\label{eq:MRR}
	MRR = \frac{1}{|S^{Test}|} \sum_{s \in S^{Test}} \frac{1}{Rank_s}
\end{equation}
where $Rank_s$ is the place of the correct item in session $s$.  From Eq. \eqref{eq:MRR}, we can see that, at most, the MRR reaches $1$ if the algorithm consistently predicts the item correctly by placing it first in the list and reaches $1/t$, where $t$ is the length of the recommended list when the recommender consistently predicts the item the last in the list. The MAP, on the other hand, evaluates the predicted list up to a specific cut-off point $N$. MAP$@N$ does not consider the order of the list. However, it checks if the correct item is present within the first $N$ items in the list.
\begin{equation}
	\label{eq:MAP}
	MAP@N = \frac{1}{|S^{Test}|} \sum_{s \in S^{Test}} \frac{Top(s,N)}{N}
\end{equation}
where $Top(s,N)$ is a function that returns $1$ if the correct item for session $s$ is in the top $N$ items in the predicted list and returns $0$ otherwise. In this experiment, we use the following values for $N$: 1, 3, 5 and 10.

\subsection{Implementation}
The experimental framework, competing methods, data preprocessing, and experimental evaluation are implemented using python with standard numerical and data manipulation libraries.

Several specialized software libraries can be used for solving the optimization problem, which is the most crucial step in our proposed method. Interior Point OPTimizer (IPOPT) \cite{wachter2006IPOPT} is a stable and well-tested advanced nonlinear programming (NLP) solver based on an interior-point filter with a line-search algorithm for large-scale nonlinear optimization. Efficient and effective linear solvers are required by IPOPT and are essential for solving the optimization problem. Therefore, The solver MA57, part of the HSL package \cite{hsl2007collection}, was used in the experiment.

IPOPT NLP solver requires, at each iteration, essential information about the optimization problem in order to solve and proceed iteratively toward the solution. At each step, the following information is required given the current values of the unknowns:
\begin{itemize}
	\item The objective function as defined in Eq. \eqref{eq:ff}.
	\item  The gradient of the objective function (vector of first derivatives with respect to all unknowns) as defined in Eq. \eqref{eq:gfi}.
	\item  The Hessian of the objective function (matrix of all second derivatives). This matrix is very large. However, the library offers the possibility of approximating the Hessian numerically. This option has been used in order to avoid large memory consumption.
\end{itemize}

\subsection{Competing Methods}
To assess the effectiveness of the proposed method, we compare it with a set of commonly used baselines and state-of-the-art methods:

\begin{itemize}
	\item Random: Even though this is not a recommendation method, it provides insight into the lowest acceptable value that other methods must achieve on this dataset.
	\item Items Popularity (I-POP): A naive baseline predictor that consistently recommends a list of the most popular items in the training set without considering user actions or similarities. Regardless of its simplicity, it often provides a stable baseline in sequence-aware recommender systems.
	\item Items Click-out Popularity (IC-POP): Similar to I-POP, however, only clicked-out items are considered when determining the item's popularity.
	\item Items Metadata $K$-Nearest Neighbors (IM-KNN): An implementation of a content-based filtering method that predicts the current user's next action based on the $k$-most similar neighbors to the previously clicked item. The similarity between the items is determined by the cosine similarity of the item's metadata.
	\item Items Co-occurrence $K$-Nearest Neighbors (IC-KNN): An implementation of an item-based collaborative filtering method that predicts the current user's next action based on the $k$-most similar neighbors to the previously clicked item. The similarity between the items is determined by the number of co-occurrence between them in sessions, calculated by cosine similarity.
	\item Logistic Regression (LR) \cite{adamczak2020session} : A method that predicts whether the item is clicked-out or not (i.e., binary classification). The method requires the extraction of specific features. We adopted the same problem formalization and feature selection as in \cite{adamczak2020session}.
	\item MLP Regressor (MLP): Multilayer perceptron is a neural network model that trains using backpropagation with no activation function in the output layer. We used the same features as in the LR model.
\end{itemize}  

\subsection{Parameters Settings and Tuning}
For the proposed method, several adjustable parameters require tuning. The number of dimensions $D$, the regularization coefficient $\lambda$, and the constant $\alpha$. In this experiment, we used grid search to determine those parameters. The number of dimensions $D$ is tried with $\{5, 10, 20\}$.  The regularization $\lambda$ is selected from the set $\{0.1, 0.01\}$. Finally, the constant $\alpha$ is chosen from $\{1, 2, 3\}$.

An important issue when fitting our model to data is the initialization step. Since the objective function is non-convex, only local minima can be found, and consequently, the initial values of the unknowns significantly affect the quality of the solution. In particular, initialization with the same values, such as setting all coordinates to 0, causes the gradient in Eq. \eqref{eq:gfi} to vanish and the optimizer to stop immediately. Hence, a random initialization is essential to avoid the trivial solution where all items are assigned the same position.

\subsection{Experimental Results}

\begin{figure*}[!t]
	\centering
	\includegraphics[height=0.23\textwidth]{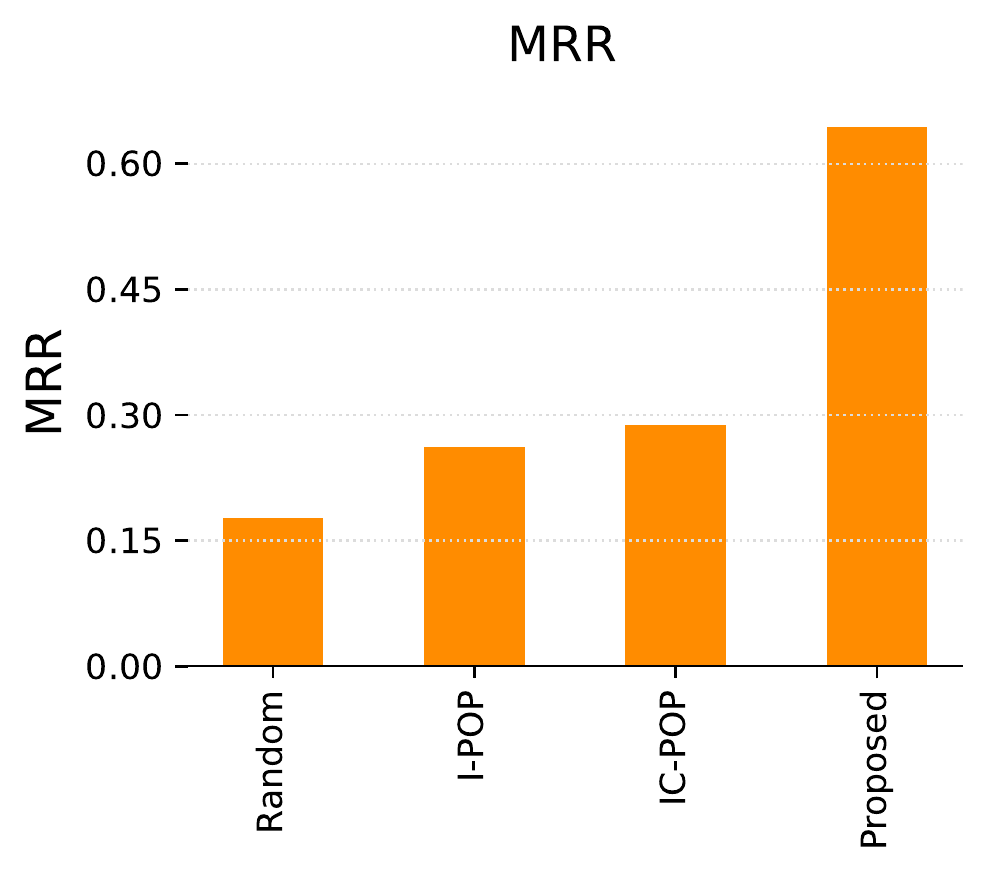}
	\quad
	\includegraphics[height=0.23\textwidth]{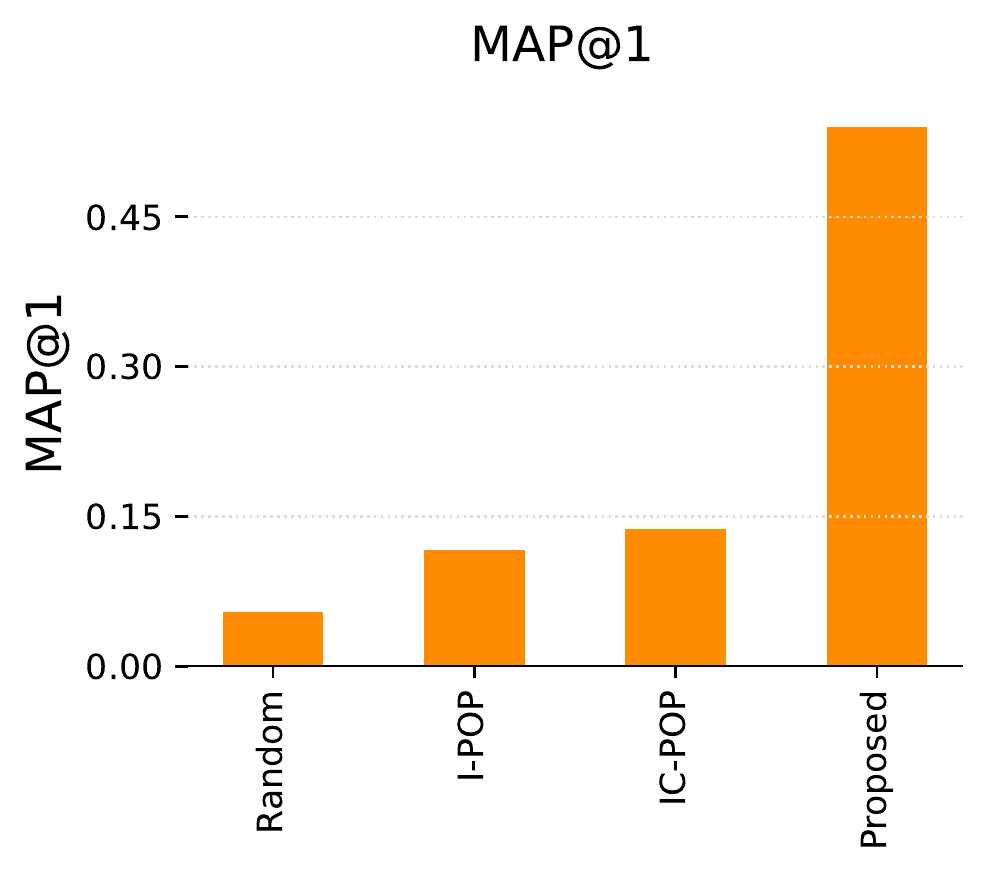}
	\quad
	\includegraphics[height=0.23\textwidth]{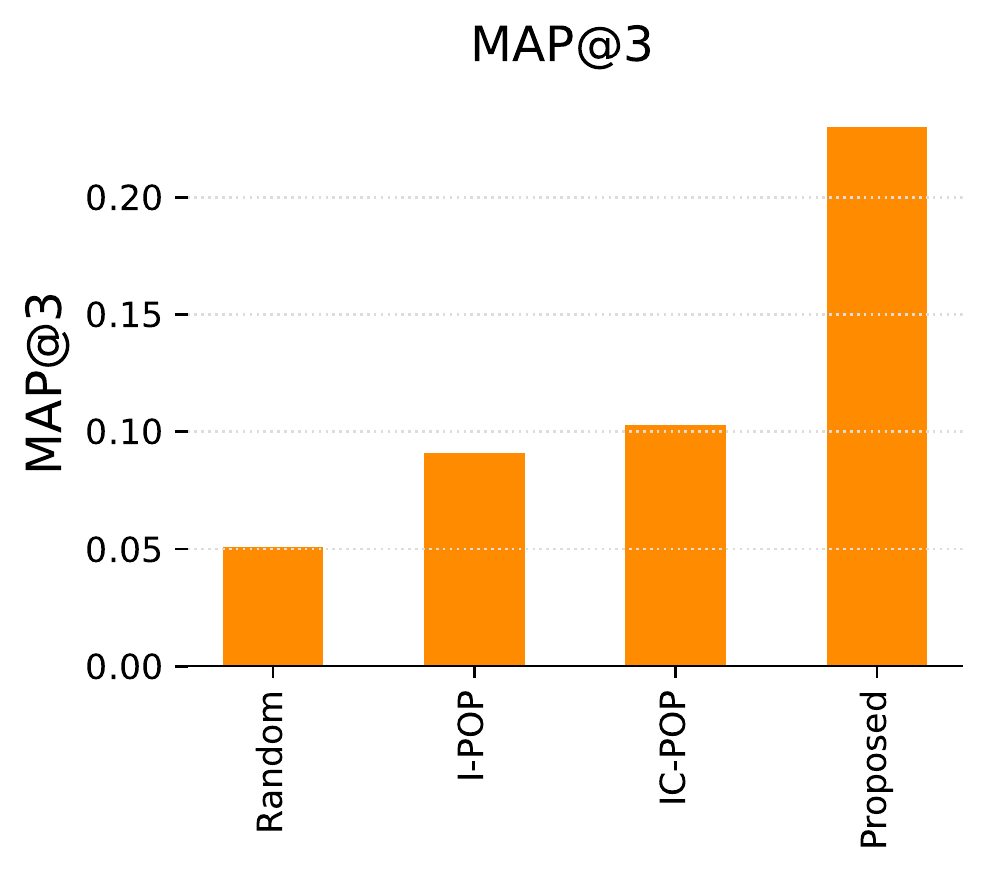}
	
	\includegraphics[height=0.23\textwidth]{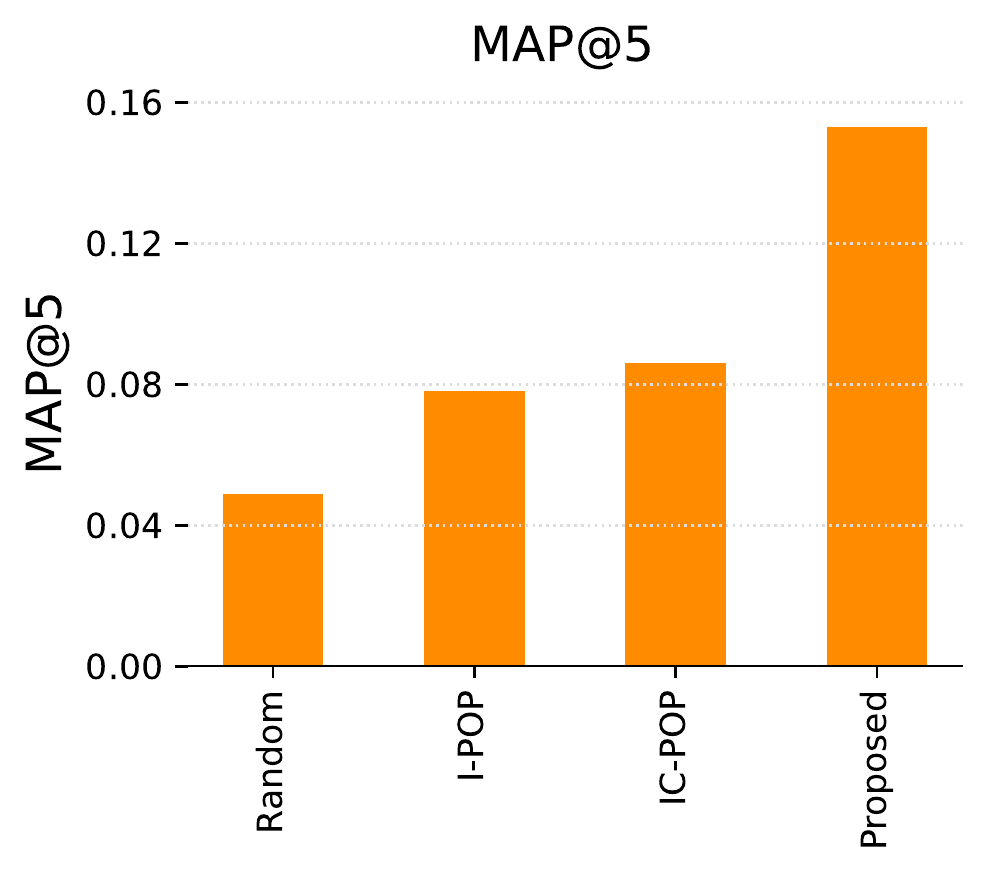}
	\quad
	\includegraphics[height=0.23\textwidth]{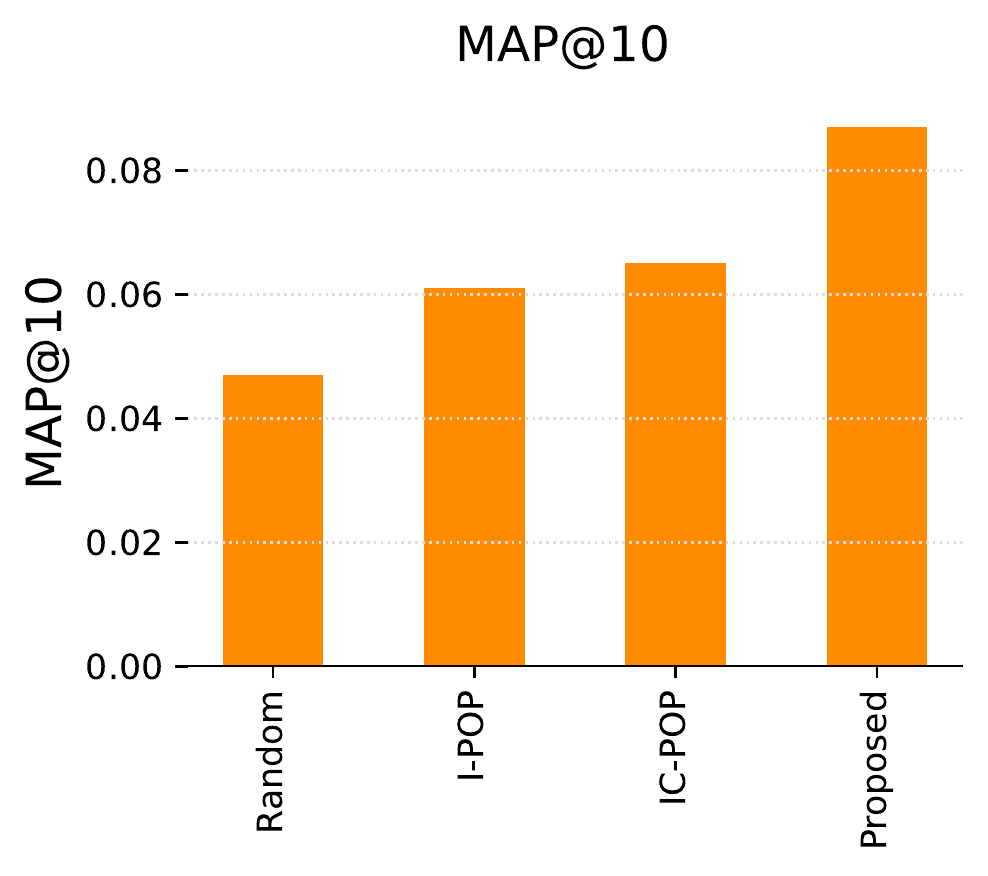}
	\caption{Comparison of the performance of the proposed approach against baseline methods on the Trivago dataset.}
	\label{fig:res-baseline}
\end{figure*}

\begin{figure*}[!t]
	\centering
	\includegraphics[height=0.23\textwidth]{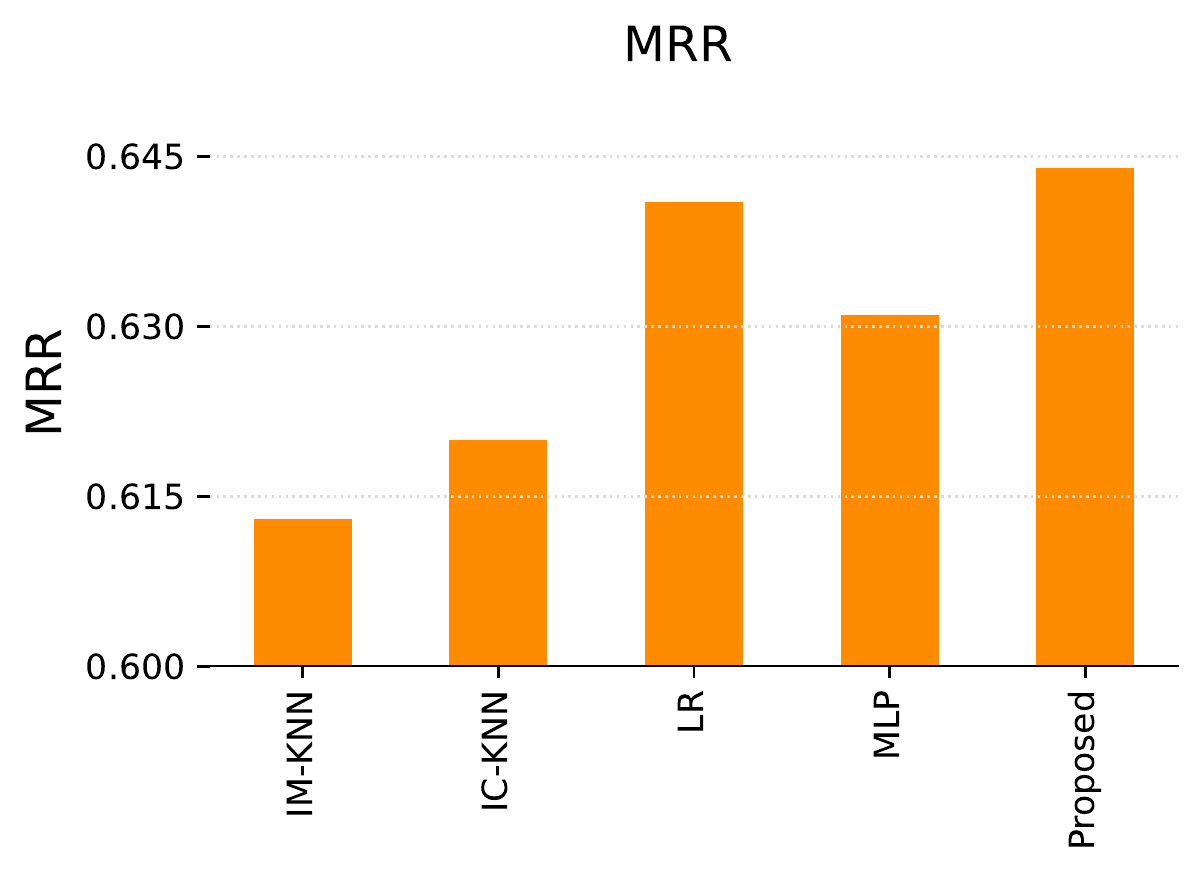}
	\quad
	\includegraphics[height=0.23\textwidth]{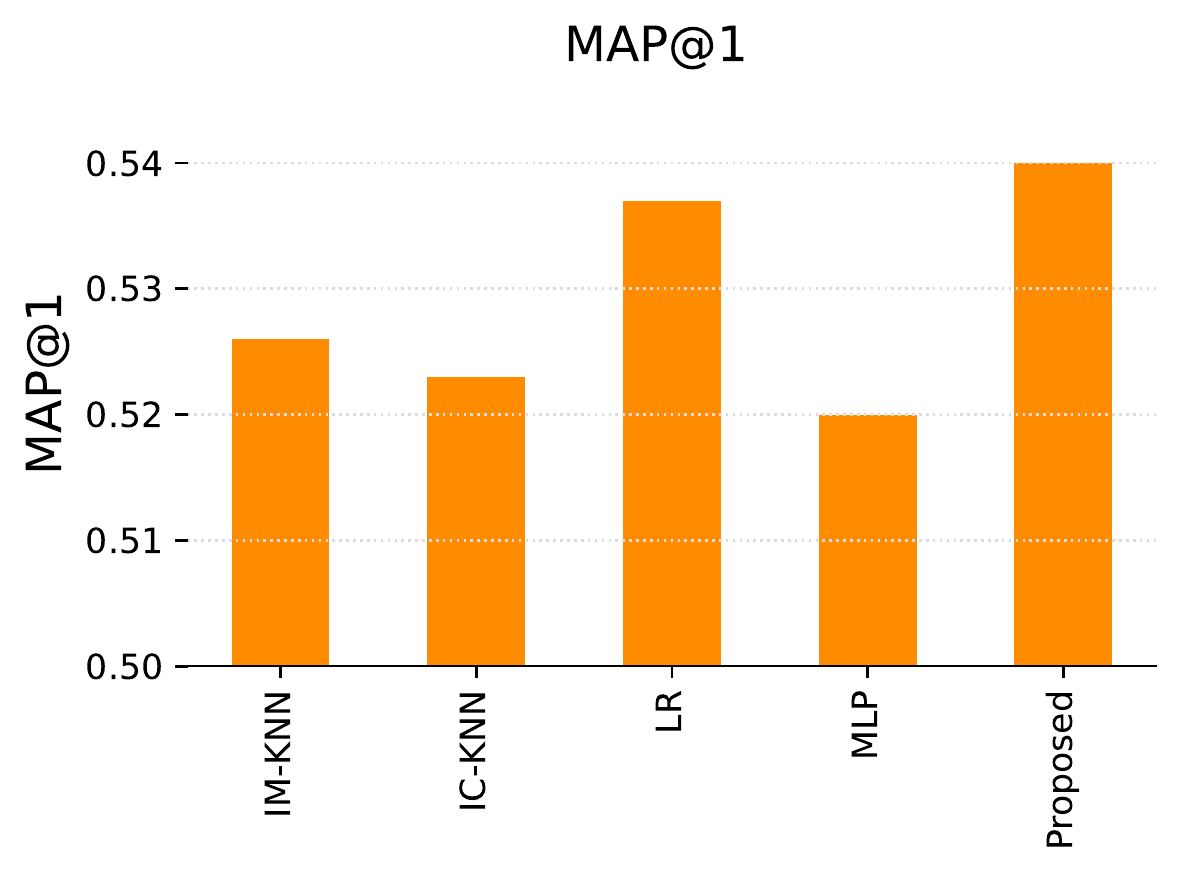}
	\quad
	\includegraphics[height=0.23\textwidth]{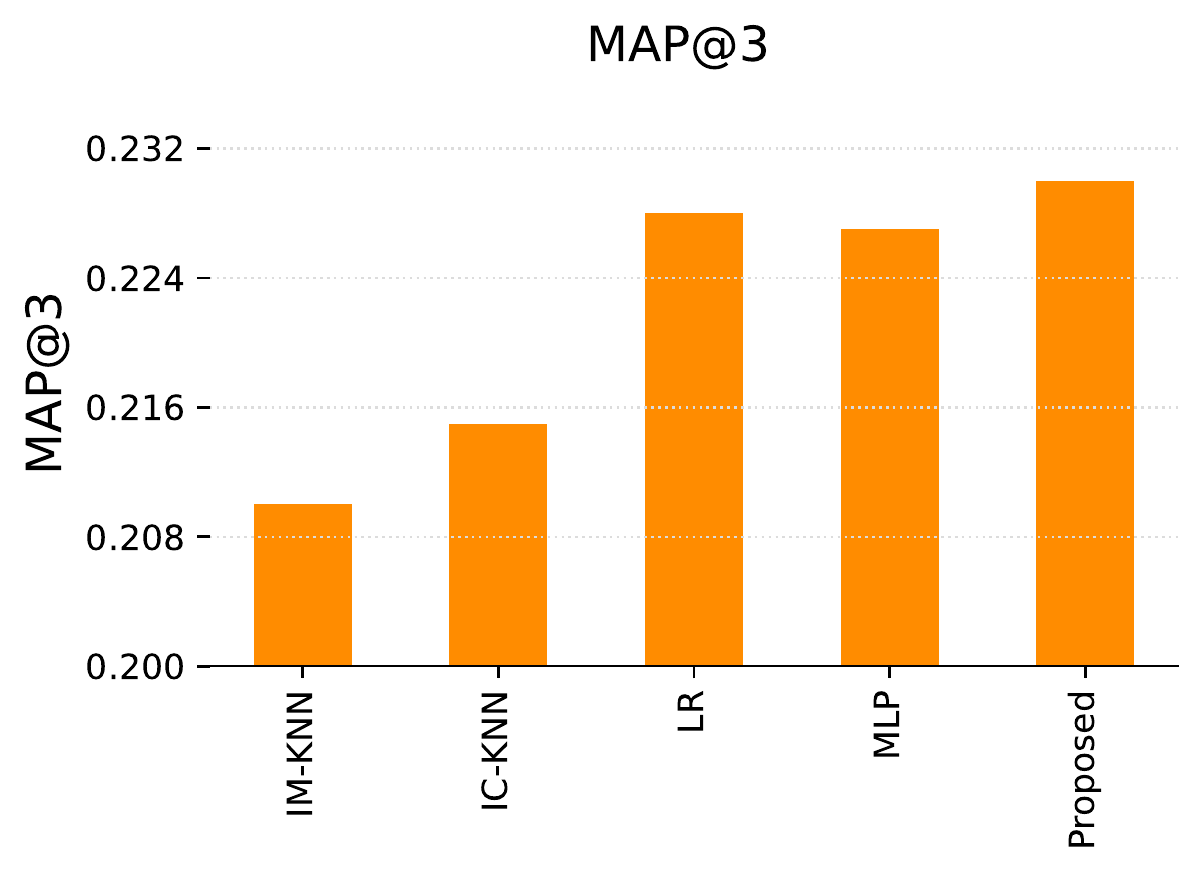}
	
	\includegraphics[height=0.23\textwidth]{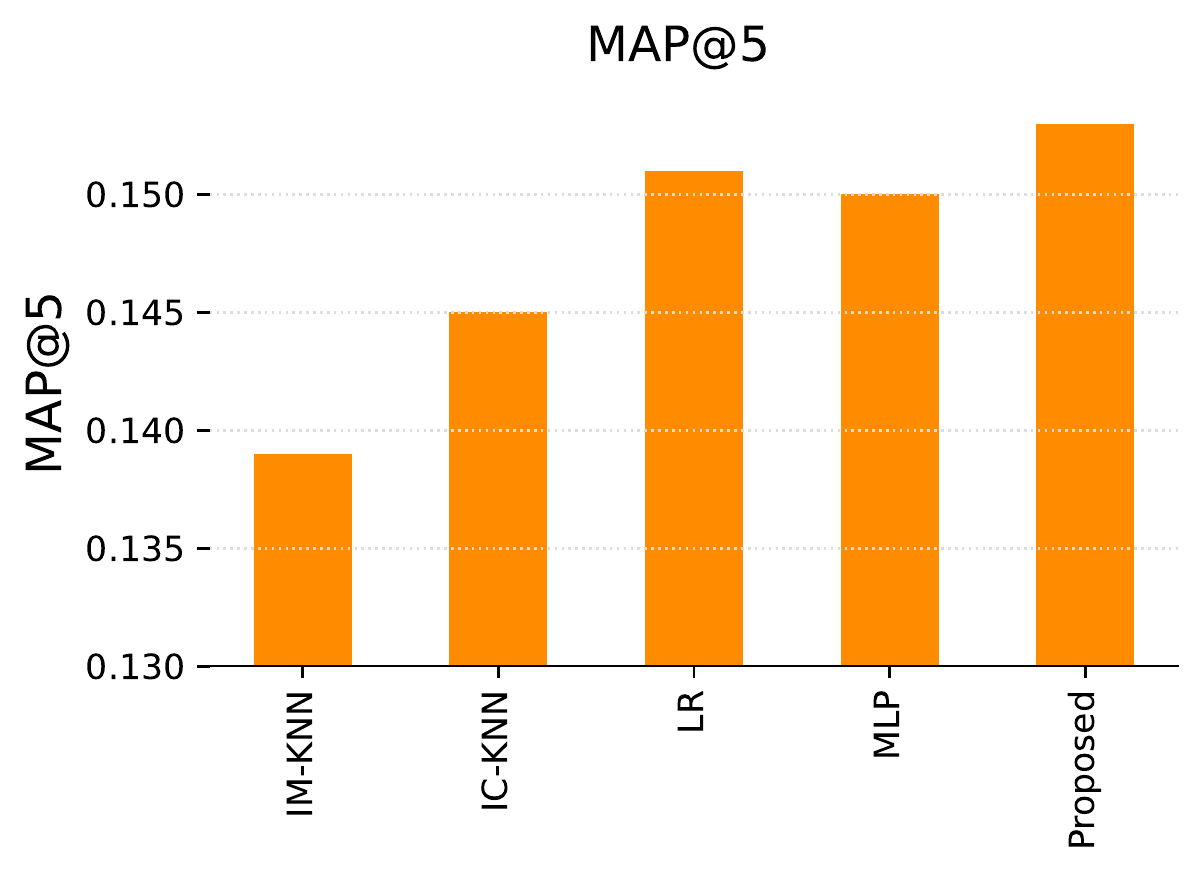}
	\quad  
	\includegraphics[height=0.23\textwidth]{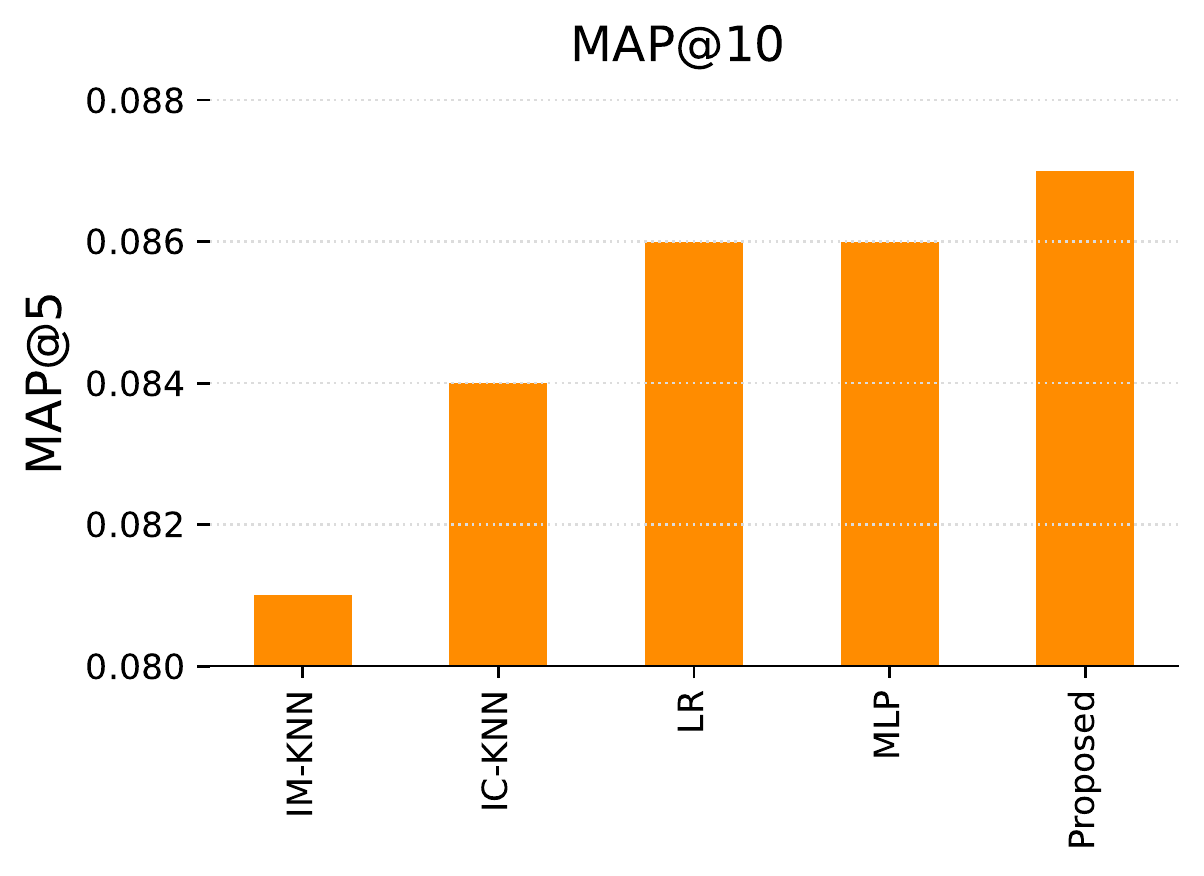}
	
	\caption{Comparison of the performance of the proposed approach against state-of-the-art methods on the Trivago dataset.}
	\label{fig:res-competing}
\end{figure*}

Table \ref{table:results} summarizes the results in terms of MRR and MAP. The results are also displayed in Figure \ref{fig:res-baseline} for the case baseline methods and Figure \ref{fig:res-competing} for state-of-the-art methods. The results show that the proposed method produces better scores than other competing methods under all considered metrics.

As expected, popularity-based methods (I-POP and IC-POP) score weak results as they rely solely on an item's popularity and ignore the user features. In other words, those methods are expected not to achieve well as they do not produce personalized recommendations.    

Interestingly, the proposed method performed better not only against  KNN-based methods (IM-KNN and IC-KNN), which rely on similarity measures without requiring feature extractions, but also against feature-based methods (MLP and LR), which require considerable effort in the feature engineering process. This is particularly important given that the Trivago dataset is rich in information, which may not be true for all datasets.

Since the proposed approach does not involve the additional semantic information available in Trivago datasets, we expect it to perform very well on datasets with limited information. 

\begin{table}[!t]
	\centering
	\caption{Performance results obtained on the Trivago dataset.}
	\label{table:results}
	\begin{tabular}{lccccc}
		\toprule
		Method   & MRR            & MAP@1          & MAP@3          & MAP@5          & MAP@10         \\ \midrule
		Random   & 0.177          & 0.054          & 0.051          & 0.049          & 0.047          \\ \midrule
		I-POP    & 0.262          & 0.116          & 0.091          & 0.078          & 0.061          \\  \midrule
		IC-POP   & 0.288          & 0.137          & 0.103          & 0.086          & 0.065          \\  \midrule
		IM-KNN   & 0.613          & 0.526          & 0.210          & 0.139          & 0.081          \\  \midrule
		IC-KNN   & 0.620          & 0.523          & 0.215          & 0.145          & 0.084          \\  \midrule
		LR       & 0.641          & 0.537          & 0.228          & 0.151          & 0.086          \\  \midrule
		MLP  & 0.631          & 0.520          & 0.227          & 0.150          & 0.086          \\  \midrule
		Proposed & \textbf{0.644} & \textbf{0.540} & \textbf{0.230} & \textbf{0.153} & \textbf{0.087} \\ 
		\bottomrule
	\end{tabular}
\end{table}

\section{Conclusion and Future Work} 
\label{sec:conclusion}
This paper demonstrates the potential of using complex networks with underlying similarity-popularity models for a sequence-aware recommendation system. The experimental results show that the proposed method performs better than state-of-the-art methods in terms of MRR and MAP. Furthermore, the proposed method does not rely on feature extraction, which makes it suitable for generalization to datasets with limited information (similar to KNN-based approaches) while providing better results. 

In future work, we propose testing the method on other datasets from various domains to confirm its generalization capacity. Moreover, additional similarity-popularity models other than the hidden metric space model can be explored. Such models range from basic similarity-popularity dot-product to more recent and sophisticated models found in the literature. In this work, once the items are embedded, we use the $k$-nearest neighbors technique to capture the user's current session and predict the next item. However, it is possible to use more advanced techniques such as deep neural networks, namely the LSTM or GRU variants of recurrent neural networks, by feeding the network the user's current sequence of actions and item embeddings as input and then training it to predict the next item.

\section*{Acknowledgment}
This research work is supported by the Research Center, CCIS, King Saud University, Riyadh, Saudi Arabia.

\IEEEtriggeratref{20}
\bibliographystyle{IEEEtran}

\end{document}